\documentclass[aps,epsfig,showpacs,pre,10pt]{revtex4-1}
\usepackage{graphicx}
\usepackage{graphics}
\usepackage{epsfig}
\begin{document}
\title{The probability analysis of opening of DNA}
\author{Shikha Srivastava$^1$ and Navin Singh$^2$}
\affiliation{$^1$Department of Physics, Banaras Hindu University, Varanasi 221 005, India \\
$^2$ Department of Physics, Birla Institute of Technology \& Science Vidya Vihar, Pilani, 333 031, 
Rajasthan, India}
\email{knavins@gmail.com}
\date{\today}

\begin{abstract}
We have studied  the separation of a double stranded DNA (dsDNA), which is driven 
either by the temperature or force. By monitoring the probability of opening 
of entire base pairs along the chain, we show that the opening of a dsDNA depends not only 
on the sequence but also on the constraints on the chain in the experimental setups. Our 
results clearly demonstrate that the force induced melting of dsDNA, whose one of 
the ends is constrained, is significantly different from the thermal melting, when 
both ends free.
\end{abstract}
\maketitle

\section{Introduction}
Until recently, information about the inter- and intra- molecular forces involved in 
the stability of the double stranded DNA (dsDNA) was obtained {\it in vitro} through
indirect physical and thermodynamic measurements like nuclear magnetic resonance 
spectroscopy, light scattering, crystallography, differential scanning calorimetry 
etc. \cite{wartell}. Such information is needed to understand two key biological 
processes, {\it i.e.}, replication and transcription, where a dsDNA is required to 
separate (fully or partially) into two single stranded DNA (ssDNA) \cite{watson}. 
It is believed that the stability of dsDNA is the result of hydrogen bonding between 
bases, backbone conformational constraints, electrostatic interactions and the coordination
of water molecules \cite{boian}. In recent years, single molecule force spectroscopy 
(SMFS) experiments using optical tweezers, atomic force microscope etc.,
have directly measured these forces and provided unprecedented insight into 
the mechanism involved in the process of DNA separation and its stability
\cite{ubm,smith,prentiss_prl,prentiss_pnas,kumar}. 

So far, there does not exist any model that accurately describes the force-induced melting 
behavior of dsDNA as a function of nucleotide sequence. Simple models, e.g, Poland 
Scheraga (PS) model \cite{poland} or Peyrard-Bishop-Dauxious (PBD) model 
\cite{pbd1,pbd2,saul} have described some of the essential macroscopic features of 
the meting profile of dsDNA quite effectively and predicted that the force-induced 
melting transition is a first order transition \cite{somen,nelson,mukamel}. 
However, semi-microscopic information about the opening such as whether a dsDNA 
opens from the end or interior of the chain, distribution of partially opened regions 
in the form of bubbles in the chain \cite{saul,kgy_epl,shikha, cocco,ns_epje,rajeev}, 
are  some of the intriguing issues in these studies. Moreover, in all SMFS experiments, 
the experimental setup puts an extra constraint on the ends of the chain, which makes 
force-induced melting different from the thermal melting. These 
constraints do not matter in the thermodynamic limit. However, all SMFS experiments 
involve a finite length of the strand and hence these constraints may affect the melting 
profile of the chain \cite{prentiss_pnas_2008,garima,ns_pre}. Singh {\it et al.} 
considered the self-avoiding walk \cite{sk_jpa} model of dsDNA \cite{ars_jcp} 
of homosequence and obtained the force-temperature diagram for a small chain length. 
It was shown that constraining the end, affects the melting profile significantly. 
The aim of this manuscript is to study the effects of these constraints on the 
melting profile of heterosequence of finite length single molecule experiments 
and provide semi microscopic information about the formation of bubbles in 
the form of partially opened regions during the opening of a dsDNA. More precisely, 
by employing the probability analysis of opening of individual bases, we delineate 
the mechanism involved in the separation of a dsDNA. For this, we adopt the PBD model, 
\cite{pbd1,saul} which has been discussed in section II. In this section, 
we also introduce the method to calculate the melting profile and probability
analysis to monitor the opening of individual base pairs (bps). Melting profiles of two 
different sequences \cite{zoli}, whose end(s) is(are) constrained by the experimental setups are 
discussed in section III. The method developed in section II, has been extended
to study the force induced melting of dsDNA in section IV. The paper ends 
with a brief discussion in section V.

\section{Model and Method}
\label{method}
\begin{figure*}[ht]
\includegraphics[height=1.5in,width=5.in]{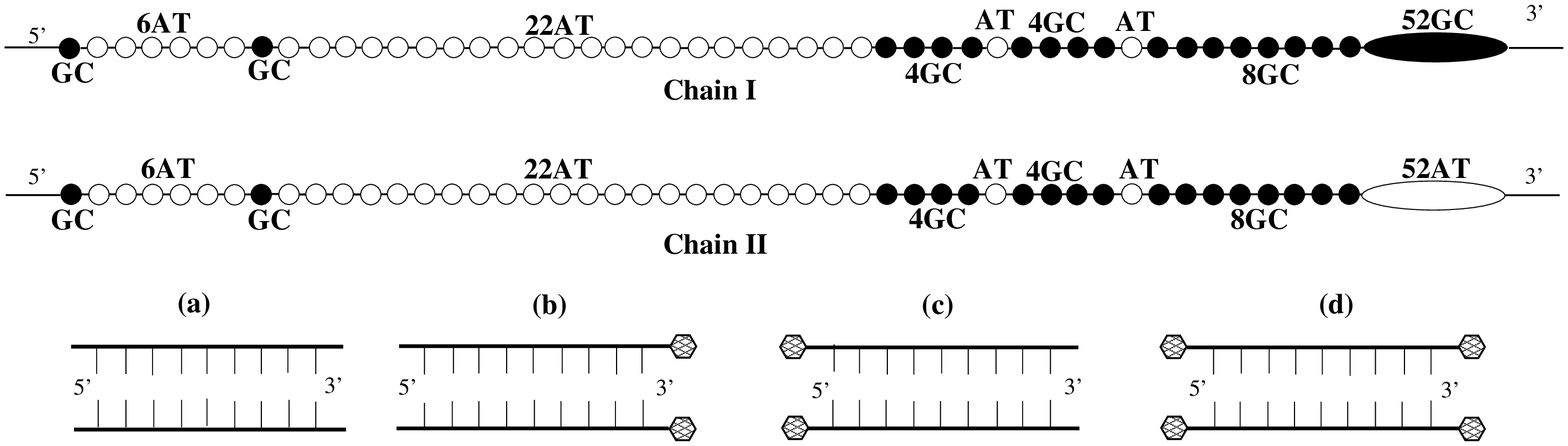}
\caption{\small Schematic representations of  two sequences (I \& II) with 
various possibilities of constrains imposed on the end(s) have been shown. 
The polygon attached with the end(s) show that the end(s) is (are) 
constrained by the experimental setup.}
\label{fig1}
\end{figure*}
In this section, we briefly discuss the basic features of the PBD model, 
which considers the stretching between corresponding bases only.
Unlike the PS model \cite{poland} which is based on the two state model 
(bound segment or unbound segment), the PBD model \cite{pbd1,saul} includes 
intermediate state because the stretching is a continuously varying variable. 
Here, we ignore the helecoidal structure \cite{cocco} of dsDNA and properties associated 
with it and focus only on the stretching of hydrogen bonds, which are represented by 
continuous variables $y_i$ ($i=1,2...... N$, where $N$ is the length of the chain).  
The sequence dependence in the model Hamiltonian can be introduced through the
potential energy term appearing in the Hamiltonian. The nonlinear term which corresponds
to the stacking energy between consecutive base pairs along the strand, mimics the long
range effects along the dsDNA strand.
The model Hamiltonian is \cite{pbd1},
\begin{equation}
\label{eqn1}
H = \sum_{i=1}^N\left[\frac{p_i^2}{2m} + V(y_i) + W(y_i,y_{i+1})\right]
\end{equation}
where $p_i = m${{\.y}$_i$}, represents the momentum part of the Hamiltonian and $m$ is 
the reduced mass of a base pair (taken to be the same for both A-T and  G-C base pairs here). 
The hydrogen bond interaction between two bases in the $i^{\rm th}$ pair is represented by, 
the Morse potential, $$V(y_i) = D_i(e^{-a_iy_i} - 1)^2,$$ where $D_i$ represents the potential 
depth, roughly equal to the bond energy and $a_i$ represents the inverse of the width
of the potential well. The stacking interaction between two consecutive base pairs along 
the chain is represented by, 
$$W(y_i,y_{i+1}) = \frac{k}{2}(y_i - y_{i+1})^2[1 + \rho e^{-b(y_i + y_{i+1})}]$$
in the Hamiltonian, where $k$ represents the single strand elasticity, $\rho$ represents the 
anharmonicity in the strand elasticity and $b$ represents its range. 
The canonical partition function of the system can be written as
\begin{equation}
\label{eqn2}
Z = \int \prod_{i=1}^{N}\left\{dy_idp_i\exp[-\beta H]\right\} = Z_pZ_c
\end{equation}
where $Z_p$ and $Z_c$ are the momentum part and configurational part of the partition function 
respectively. Since the momentum part is decoupled in the integration, it can be integrated 
out as a simple Gaussian integral, which contributes a factor $(2\pi m k_BT)^{N/2}$ to the 
partition function \cite{pbd1}. The configurational part of the partition function, $Z_c$, 
is
\begin{equation}
\label{eqn3}
Z_c =\int_{-\infty}^{\infty}\prod_{i = 1}^{N} dy_i  \exp[-\beta H(y_i,y_{i+1})].
\end{equation}
For the homo sequence, one can evaluate the partition function by the transfer integral 
(TI) method with periodic boundary condition. In case of heterosequence the integration
in the configurational partition function can be carried out numerically with the help 
of matrix multiplication method \cite{chen,campa,ns_epje}. Once the limits of integration 
have been chosen, the task is to discretized the space so that the integral can be 
evaluated numerically. Here, we fix the lower and upper limits  of the integration to be
$-5.0 \; {\rm \AA}\; {\rm and}\; 200.0 \; {\rm \AA}$ respectively. The space is 
discretized using the Gaussian quadrature formula \cite{nrc} with a certain number of grid 
points. In our previous study \cite{ns_pre}, it was  shown  that to get precise value 
of the melting temperature ($T_m$), one has to choose large number of grid points. In this study, 
900 grid points are found to be sufficient to obtain the melting profile of the 
chain. Since all the  matrices in Eq.(3) are identical, the multiplication is straight 
forward. The free energy per base pair is obtained from the following relation,
\begin{equation}
\label{eqn4}
f(T) = -\frac{1}{2}k_B T\ln\left(2\pi m k_B T\right) - \frac{k_B T}{N}\ln Z_c
\end{equation}
The thermodynamic quantities of interest {\it e.g.} the number of intact base pairs ($\theta$) 
\cite{campa,note} and the specific heat ($C_v$) of the system may be evaluated by using the 
following relations, 
\begin{equation}
\label{eqn5}
\theta = \frac{1}{N}\sum_{i=1}^{i=N}\langle\Theta(y_0-y_i)\rangle \;\; \& \;\;
C_v(T) = -T\frac{\partial^2 f}{\partial T^2}
\end{equation}
where $\Theta(y)$ is the Heaviside step function. We set the limit, $y_0$ equal 
to 1${\rm \AA}$, above which a base pair is considered to be in the open state. 
The emphasis of present work is to evaluate the probability of opening of 
particular base pairs at a given external condition. The probability of opening of 
the $j^{\rm th}$ pair in the sequence is defined as \cite{rapti_pre},
\begin{eqnarray}
\label{eqn6}
P_j &=& \frac{1}{Z_c}\int_{y_0}^{\infty}dy_j \exp[-\beta H(y_{j},y_{j+1})]Z_j
\end{eqnarray}
where 
\begin{equation}
Z_c = \int_{-\infty}^{\infty}\exp\left[-\beta \frac{V(y_1)}{2}\right]\prod_{i = 1}^{N}  
dy_i  \exp[-\beta H(y_i,y_{i+1})] 
\exp\left[-\beta \frac{V(y_N)}{2}\right]
\end{equation}
is the partition function of the chain for the open boundary condition and
\begin{equation}
Z_j = \int_{-\infty}^{\infty}\prod_{{i=1},{i\neq j}}^{N} dy_i \exp[\beta H(y_i,y_{i+1})]
\end{equation}

\section{Thermal melting}
\begin{figure}[b]
\label{fig2}
\vspace{1cm}
\includegraphics[scale=0.5]{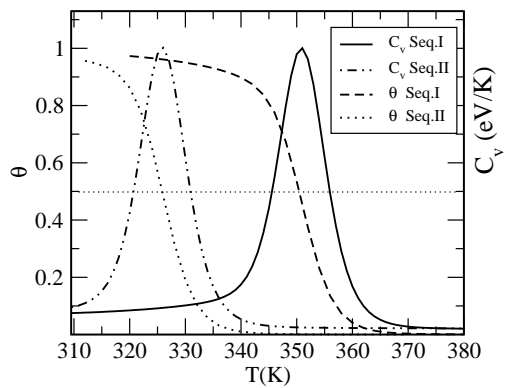}
\caption{\small Variation of number of intact base pairs and specific heat with
temperature for sequence I and II. The peak of specific heat coincides with the
value where half of the base pairs are open. Here $\it {y-axis}$ is scaled by 
peak value of $C_v$  } 
\vspace{0.5cm}
\end{figure}

We first investigate the thermal melting of dsDNA in the absence of end constraints and 
applied force. The values of the relevant model parameters are chosen in such a way that 
the  melting temperature of long homogeneous sequences, is reproduced \cite{pbd1,chen}.
The values  $D_{\rm AT} = 0.05 \;  {\rm eV},\; D_{\rm GC} 
= 0.075 \; {\rm eV}, \; a_{\rm AT} = 4.2 \; {\rm \AA^{-1}},\;  a_{\rm GC} = 6.9 \; 
{\rm \AA^{-1}}, \; \rho = 2.0, \; k = 0.025 \; {\rm eV/\AA^2}, \; b = 0.35 \; {\rm \AA^{-1}}$,  
\cite{campa} are used in the partition function to obtain the thermodynamic quantities. 
After fixing the parameters, we consider the two sequences \cite{zoli}, as shown in Fig. 1,
to study the probability distribution of bubbles and the location of their initiation. It may be 
noted here that the two sequences are almost the same except that in the sequence I, half of 
the base pairs are of GC type (we call it as GC rich end) while  sequence II has AT rich end 
(Fig. 1).

\begin{figure}[t]
\label{fig3}
\includegraphics[scale=0.45]{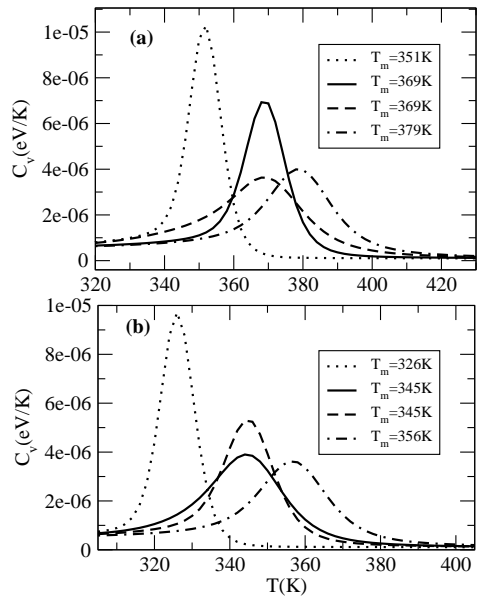}  
\caption{\small (a) Variation of specific heat with temperature for various constrains for 
sequence I. The dotted line, solid line, dashed line and dotted-dashed line correspond to 
the FIG. 1 a, b, c and d respectively. It is obvious from the plots that peak  positions 
coincides for Fig. 1b and 1c. (b) Same as Fig. 3a, but for sequence II. } 
\end{figure}
First we consider a situation, where both ends of the given sequences are free (Fig.1a). This has 
been well studied in the context of thermal melting. For short sequences, the transition is well 
understood by the two state theory and the chain opens from the end. Using Eq. (\ref{eqn2}-7), 
we obtained the melting profiles, which is  shown in Fig. 2 and  calculated the melting 
temperature $T_m$, where half of the total base pairs of a given sequence are open. It is to be 
noted that the peak in the specific heat coincides with the melting temperature and is consistent 
with earlier value obtained through path integration technique \cite{zoli}. This corresponds to a 
phase transition where the system goes from the bound state to the open state. Interestingly, 
the sequence which has GC rich end exhibits the higher melting temperature, shows that sequence I 
is more stable than the sequence II. 
\begin{figure*}[ht]
\begin{center}
\begin{tabular}{cc}
\includegraphics[height=2.8in,width=3.2in]{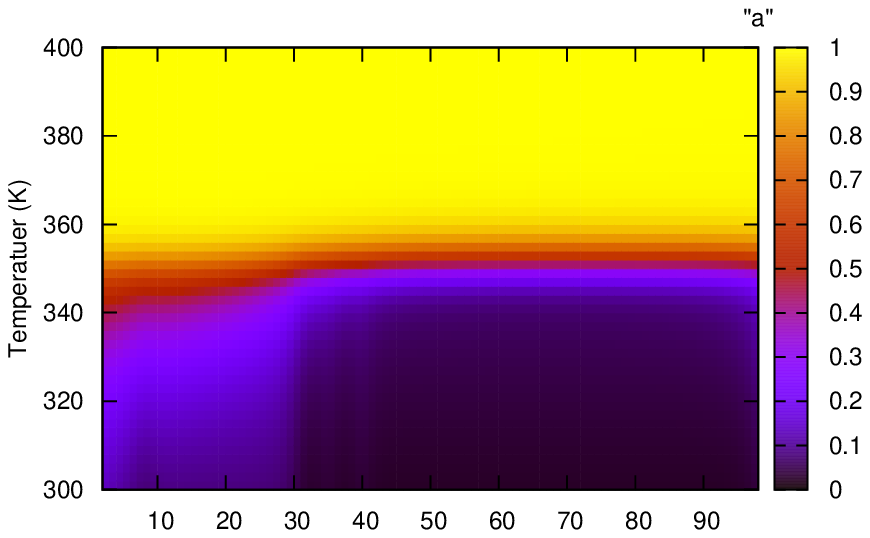} &
\includegraphics[height=2.8in,width=3.2in]{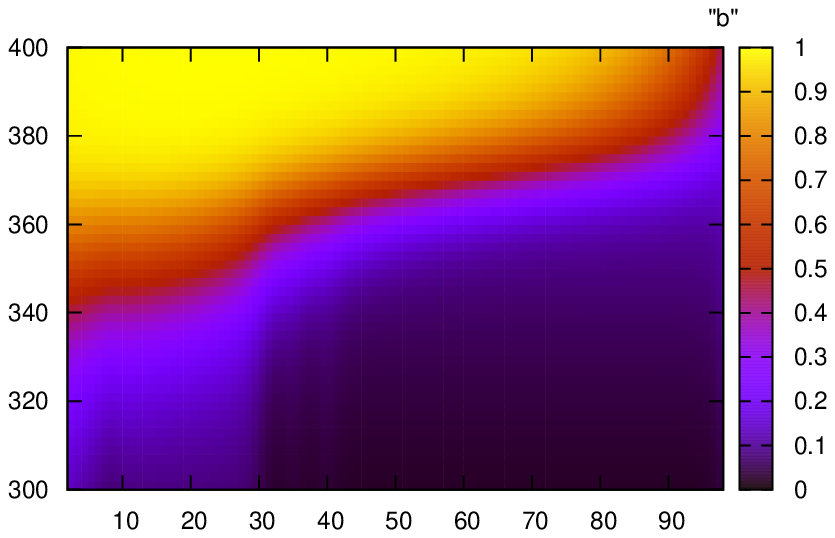} \\
\includegraphics[height=2.8in,width=3.2in]{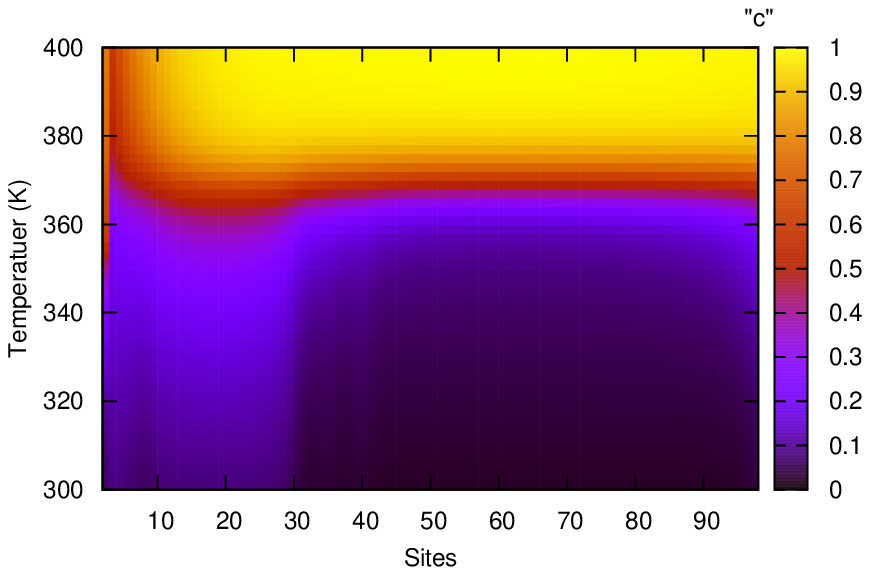} &
\includegraphics[height=2.8in,width=3.2in]{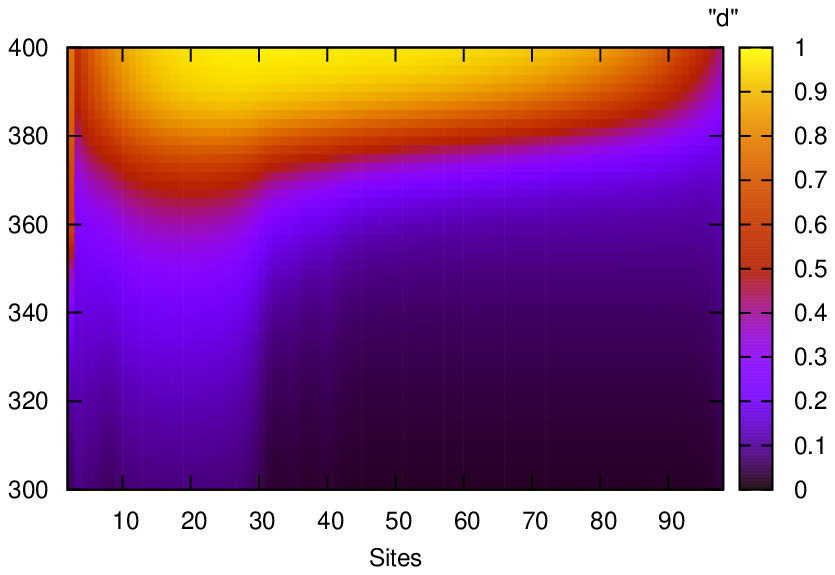} \\
\end{tabular}
\caption{\small The probability profiles of the sequence I for four different conformation as
suggested in Fig.1. There is a striking difference in the opening of the chain under four 
different conformations. The figures a, b, c \& d correspond to the four conformations as
shown in Fig 1.}
\label{fig4}
\end{center}
\end{figure*}

It may be noted that both the sequences have a block of 22 AT base pairs (starting from $31^{\rm st}$ 
position from the 5' -end), which is expected to open first and form a bubble (Fig.1). Here, our aim is 
to study the melting profile, if one of the ends, say the GC rich end (or AT rich end) is kept free 
(Fig. 1(b) or Fig. 1(c)) and, the other end is constrained. We would like to compare it with the situation 
when when GC rich end (AT rich end) is constrained, where as the other end is free. For both the cases, 
semi-microscopic information about opening of dsDNA will be obtained by monitoring the block of 22 AT 
base pairs. It should be reminded here that all SMFS experimental setups do impose such constraint on 
one end of the dsDNA while the force is applied at the other end. In order to have better insight of 
these experiments, constraining one end of dsDNA is a prerequisite for the theoretical understanding. 
During transcription, the dsDNA opens in the middle of the strand, {\it i.e.}, both ends are tied. 
Such situations can be studied, if we fix both ends of the strand as shown in Fig. 1(d) and compare its 
melting profile with (i) both ends free and (ii) only one end (GC rich end or AT rich end) 
is free. In Fig. 3 we show the plot of specific heat with temperature for both the 
sequences. It is obvious from the plots that the melting temperature remains unaffected 
whether GC rich end ($N^{\rm th}$ base pair, {\it i.e.} 3' -end) is fixed or the other 
end ($1^{\rm st}$ base pair {\it i.e.} 5' -end). One can notice that for sequence I, 
$5^{\prime}$ end of the chain is weaker than the $3^{\prime}$ -end (GC rich end). Therefore, 
the widths of the curves for the two cases are significantly different. This suggests that, 
though melting temperature of the system remains the same, the mechanism involved in the 
opening of the chain depends on the constraint imposed at the end. 

In the following, we  now  explore how the chain opens. First, we consider the probability profiles 
of the sequence I under different constraints imposed at the ends. When the chain is free at both ends, 
the probability profile (Fig. 4a) of the chain shows that the chain opens smoothly from the weaker 
segment (5' -end) to the stronger segment (3' -end). Fig. 4c is the plot for the situation when the 
chain is constrained  at the 5' -end. From the profile, it is apparent that the end effect plays an 
important role in the DNA melting for a chain of smaller length. There is a possibility of formation 
of loops in the chain near the 5' -end, which spans 22 base pairs (blue regime of the plot). 
Since the 5' -end is constrained, the loop entropy starts contributing to the system. Note that the 
GC rich segment also puts an extra constraint on the weaker segment of the chain. This means that 
the chain can open now from the GC rich end. For higher temperature range, a segment of 30-99 base 
pairs is still intact (the black region). The most interesting feature of this plot is the higher 
value of the opening probability of 3' -end as compared to its neighbour. This reveals that although 
the segment 50-100 base pairs is stronger, the chain opens from that end in case of thermal denaturation.
As temperature increases, though the GC rich end opens up the probability of opening of the AT sequence,
which forms a loop of 30 base pairs dominates the melting. From this plot the melting temperature of 
the chain can also be predicted. 

In contrast, similar features have  not been observed when we constrain the GC rich end, {\it i.e.} 
the 3' -end. Since this end has the strongest segment (GC base pairs), the chain opens from the other 
end. Although, the melting temperatures of the chain for both the cases (constrained at either of the 
ends) are found to be the  same, the nature of opening is significantly different. In this case, the
strands open gradually from the 5' -end with an interface at $30^{\rm th}$ base pair, which demarcates 
the two regions of the chain; the weaker segment and the stronger segment. It is also reflected 
that 50\% of the chain open around 370 K which is consistent with the melting temperature of the chain.

When both ends are constrained, the opening of the chain is clearly different from the above two cases. 
It has relevance in understanding the process like transcription.  Such constraints ensure that the end 
effects are suppressed. Hence, the chain will be denatured due to the entropic contribution from the 
bubble(s) that forms inside the chain. Again, from the profile one can note that 50\% of the base pairs 
open around 380 K. All these features simply indicate that constraining the end of dsDNA affects the 
local melting and affects the mechanism of  opening of the chain.
\vspace{0.5cm}
\begin{figure*}[ht]
\begin{center}
\begin{tabular}{cc}
\includegraphics[height=2.8in,width=3.2in]{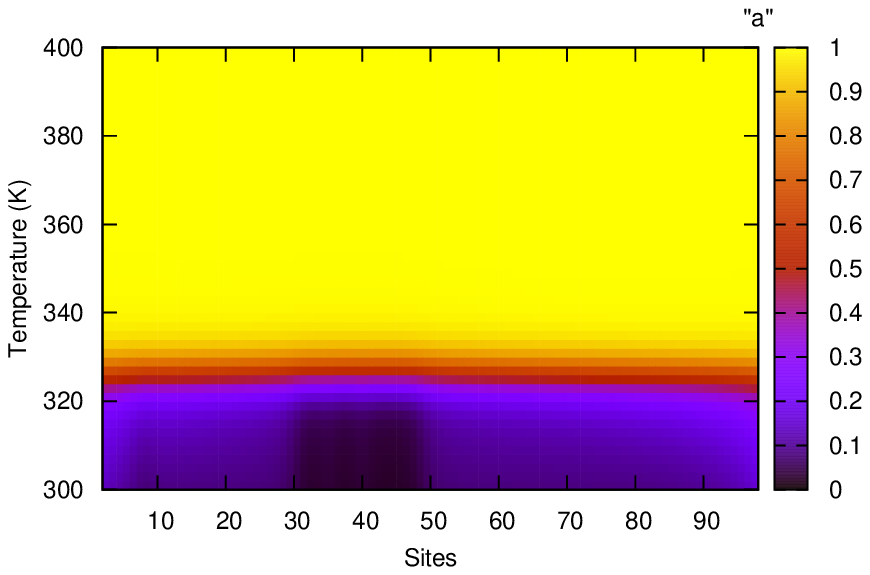} &
\includegraphics[height=2.8in,width=3.2in]{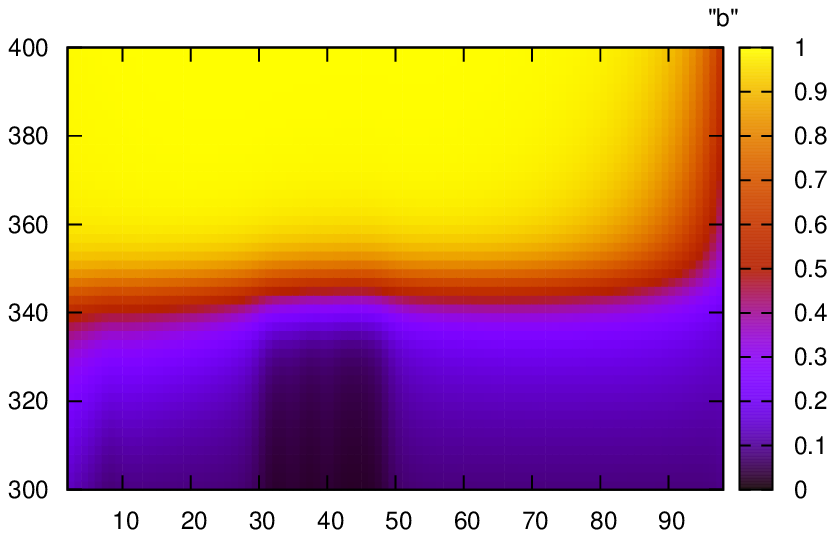} \\
\includegraphics[height=2.8in,width=3.2in]{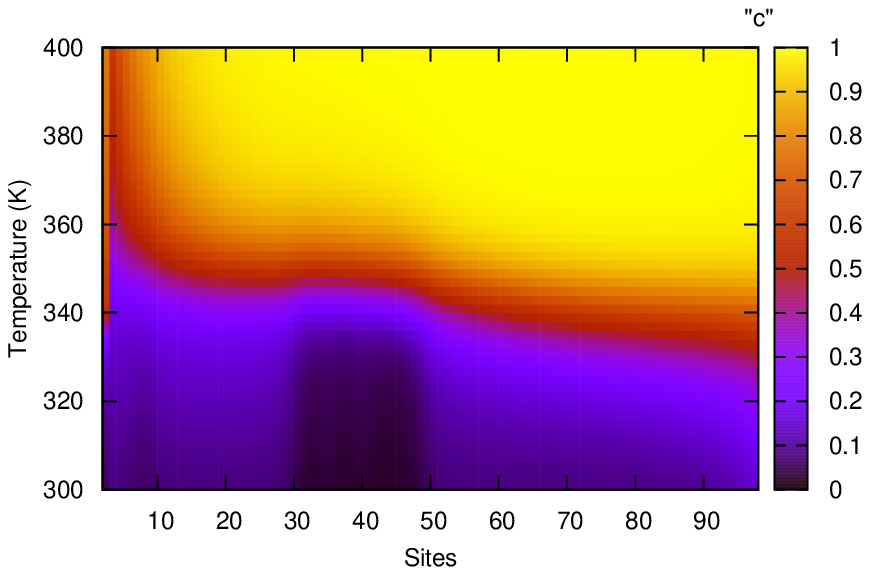} &
\includegraphics[height=2.8in,width=3.2in]{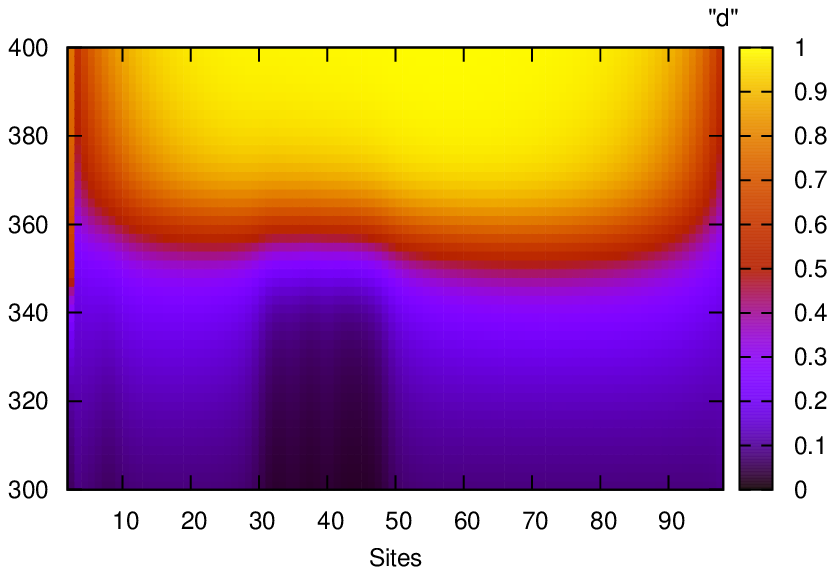} \\
\end{tabular}
\caption{\small For comparison we have plotted the density profile of sequence II.
The figure (a)-(d) corresponds to the conformations as shown in Figure 1(a)-(d). }
\label{fig5}
\end{center}
\end{figure*}

In order to have a better understanding of the role of the sequence on the opening mechanism of dsDNA, 
we have also studied the opening of sequence II. When the chain is constrained from either of the
ends, the resistance from the stronger segment which span about 20 base pairs (31-49) is clearly
visible. For the case when chain is constrained from 5' -end, chain has the two weaker segments one at 
the 3'-end and the another close to 5'-end. In addition to the resistance from the stonger segment, 
a small patch of about 5 GC base pairs puts an extra constraint, which restricts the opening from 
the 5'-end. As a result, the probability of 
opening of 3'-end is found to be larger, indicating that the chain opens from the 3'-end. The melting 
temperature of this chain is lower than the sequence I. For rest of the conformations, as observed for 
sequence I, such kind of striking difference has not been observed. This sequence has more end point 
entropy in comparison to sequence I. Thus, constraining either of the ends simply means that the 
contribution to melting from the end is being restricted (Fig 1 b \& c).  It may be noted the 
sequence II has a segment of 52 base pairs of AT at 3' -end while about 28 base pairs of AT at 5' -end. 
In the case when both ends are constrained, the chain opens from its weaker segments.

\section{Force induced transition}
We now investigate how the opening mechanism of the sequences changes when the chain is subjected to 
an external force at either ends. The modified Hamiltonian of the system under the applied force can be 
written as
\begin{eqnarray}
\label{eqn1}
H_f = H - F\cdot y_e
\end{eqnarray}
We have included a term $F\cdot y_e$ in Eq. (\ref{eqn1}), which shows that force is applied on the 
end pair. Here, our emphasis is not to calculate the critical force \cite{ns_pre}, but to focus on 
the way the chain opens by calculating the probability of opening of base pairs as a function of 
the applied force. 
\label{force}
\begin{figure*}[t]
\label{fig6}
\parbox{6.5in}{\includegraphics[height=2.8in,width=3.2in]{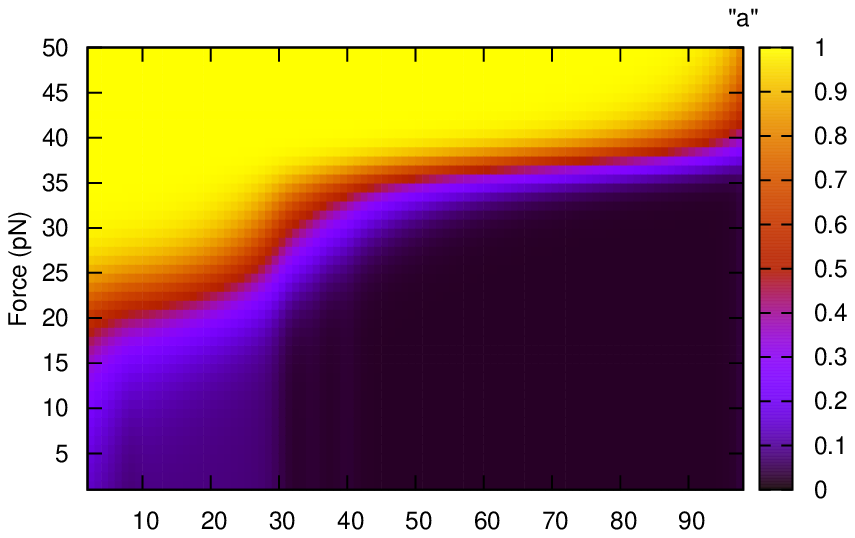}
\includegraphics[height=2.8in,width=3.2in]{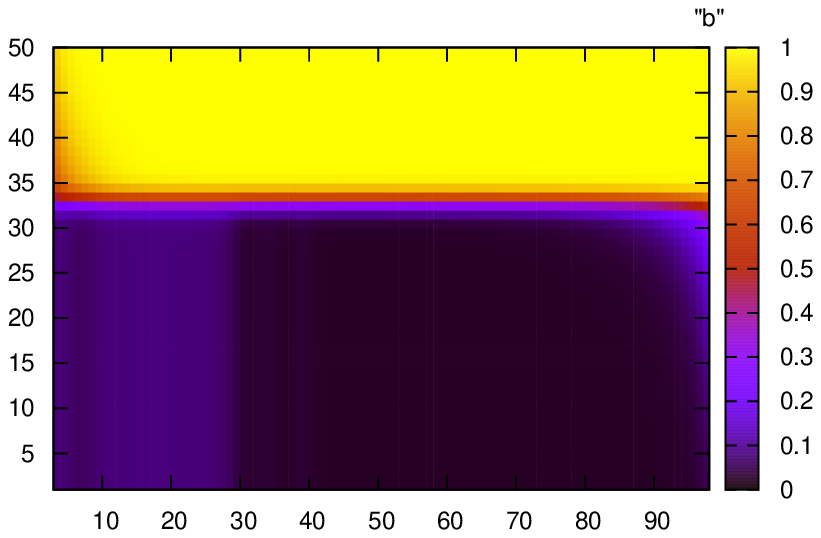}
\includegraphics[height=2.8in,width=3.2in]{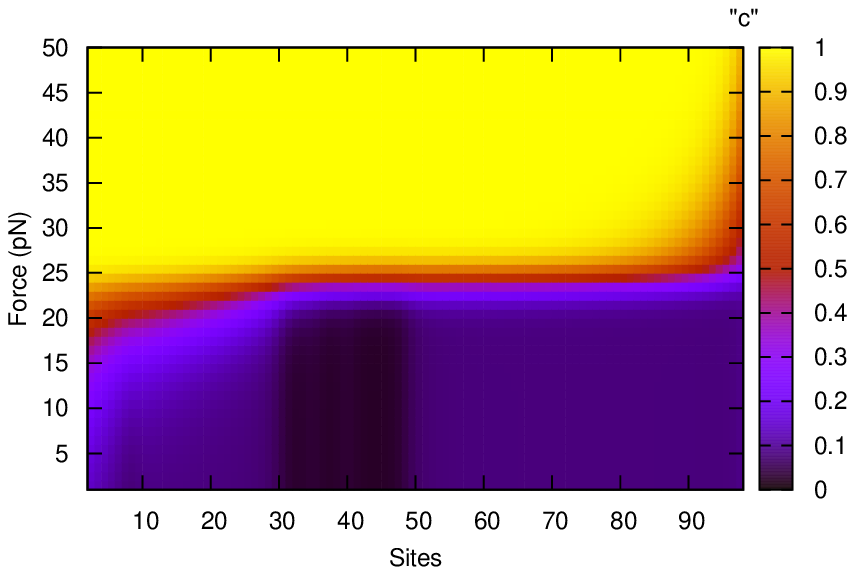}
\includegraphics[height=2.8in,width=3.2in]{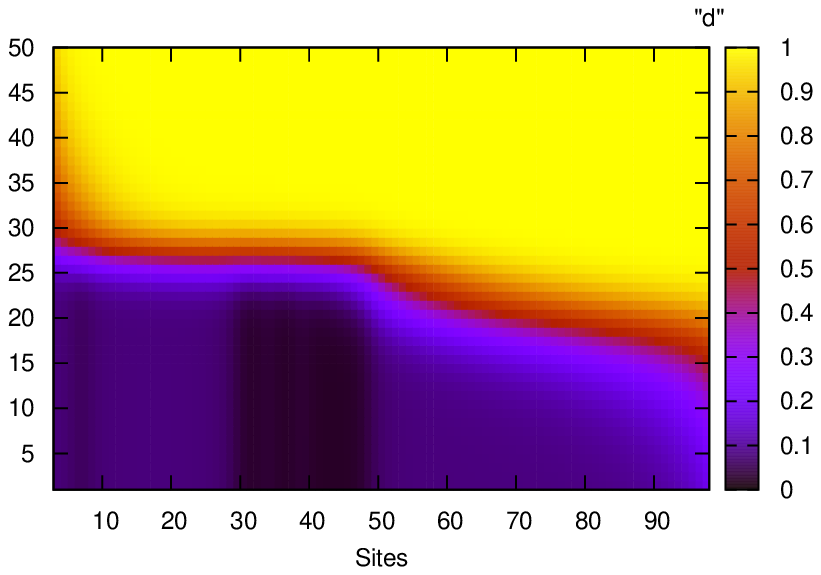}}
\caption{\small The probability profile of both the chains under the mechanical stress at
3' \& 5' -ends at T = 300 K. The figure on top left (a) is when the sequence I is constrained 
at 3'-end and force is applied at 5'-end while figure on top right (b) is for the same chain 
under reversed conditions. The nature of opening is clearly different in the two cases. 
While the melting temperature remains unaffected by the way we constrain an end, the force 
induced unzipping depends on which end is being constrained. When 3'-end is constrained the 
opening is sharp while it is smooth in the reverse case. The last two figures (c \& d) are 
for sequence II under the same conditions.}
\end{figure*}

The plots which show the influence of applied force at one end while the other end is
fixed, are shown in Fig. 6 in the form of probability profiles for sequences I and II. 
These plots clearly demonstrate that the opening of the chains under mechanical stress 
is significantly different from thermal melting. When a force is applied on the 
$1^{\rm st}$ pair (5'-end), keeping the $N^{\rm th}$ pair (3'-end) constrained, sequence I 
(Fig 6a), opens from the weaker segment and the transition from the zipped state to 
unzipped state is quite smooth. In contrast, for sequence II, although there is a force 
on the $1^{\rm st}$ pair, there is an entropic contribution because of the formation of 
a bubble (about 60 base pairs) at the other end (Fig 6c). The critical forces for the two 
sequences are found to differ by 10 pN, as sequence II is weaker than the sequence I. This is because 
at 300 K, the segments containing AT pairs have a large probability of opening, and thus 
the chain opens from both ends and critical force reduces. For sequence I, when the force 
is applied at the 3'-end, keeping the 5'-end constrained (Fig 6b), one requires a bit larger 
force to open the stronger segment of the chain. One can see from the plot that the 
probability of opening of the 30 base pairs near the 5'-end is around 0.25, even though
the chain is constrained at that end. We find that the transition is quite sharp and 
the chain opens abruptly as the GC segment opens up. For sequence II, we do not observe 
similar behaviour as there are stronger segments of about 20 base pairs in between the two 
weaker segments of AT base pairs. As a result, the chain opens from the 3'-end, but the loop
containing AT base pairs reduces the unzipping force. However, the opening mechanism does 
not change because of it. This suggests that the unzipping force depends on the sequence 
of the chain as well as the on the end which has been constrained.

\section{conclusion}
In this paper, we have studied the thermal melting and the force induced unzipping 
of two sequences having GC rich end and AT rich end, respectively. The melting temperature
of these sequences only give information about half of the base pairs, which are open. 
Using PBD model, we have calculated the probability of opening of the entire sequence to 
see how the chain opens. Motivated by recent SMFS experiments, we constrained one end of 
the chain and showed that the mechanism of opening of dsDNA not only depends on the sequence 
of base pairs, but also on which end has been constrained by the experimental setup. It would 
be interesting to repeat some of these experiments by interchanging the constrained ends. 
It is also surprising to note that in the study of the force induced unzipping using PBD model, 
the bubbles do not play a significant role. This could be due to fact that in this model, 
opening of a bead is restricted to move in one dimension only and thus entropy of the loop has 
been underestimated. It would be nice to repeat these experiments by attaching  different tags 
at the weaker junction and to see how the chain opens during thermal melting and force 
induced unzipping.

\section*{Acknowledgement}
We would like to thank Sanjay Kumar for suggesting this problem and many helpful discussions on the 
subject. We also thank Y. Singh and R.R. Mishra for the critical reading of this manuscript. 
We acknowledge the financial support provided by the Department of Science and Technology and the 
University Grants Commission, India.


\begin{thebibliography}{99}
\bibitem{note}
In fact, $\theta = \theta_{\rm ext} \theta_{\rm int}$, where $\theta_{\rm ext}$
is the average fraction of strands forming double strand and $\theta_{\rm int}$ is 
the average fraction of intact bonds in the double strand. For long chains, the 
dissociation equilibrium can be neglect in view of the fact that while $\theta$ and thus 
$\theta_{\rm int}$ varies from 1 to 0, $\theta_{\rm ext}$ is still remains 1.
\end{thebibliography}
\end{document}